# Formation of soap bubbles by gas jet

M.L. Zhou, M. Li, Z.Y. Chen, J.F. Han[*], D. Liu

Key Laboratory of Radiation Physics and Technology of the Ministry of Education, Institute of Nuclear Science and Technology, Sichuan University, Chengdu, People's Republic of China

Soap bubbles can be easily generated by varies methods, while their formation process is complicated and still worth study. A model about the bubble formation process was proposed in Phys. Rev. Lett. 116, 077801 recently, and it was reported that the bubbles were formed when the gas blowing velocity was above one threshold. However, after repeating these experiments, we found the bubbles could be generated in two velocities ranges which corresponded to laminar and turbulent gas jet respectively, and the predicted threshold was only effective for turbulent gas flow. The study revealed that the bubble formation was mainly influenced by the aerodynamics of the gas jet blowing to the film, and these results will help to further understand the formation mechanism of the soap bubble as well as the interaction between gas jet and thin liquid film.

Soap bubble has been broadly applied in entertainment, education, art and other fields for its flexible formation, variable shape and colorful appearance. Soap bubbles are also deeply studied by scientist for its esoteric properties in mathematics and physics [1-6]. The minimal surface of soap bubbles [7] gave lots of inspiration to architect Frei Otto who proposed the revolutionary tensile roof structures [8], and several famous buildings such as the West German Pavilion at Expo 67 in Montreal were constructed. Although various characteristics of soap bubbles have been extensively studied, the formation process of soap bubbles is complicated and still need deep study. According to John Davidson [9], the birth process of soap bubbles is similar to that of liquid drops. Normally the droplets are formed through the breakup of a continuous liquid jet which has been studied deeply for centuries [10-12], and the crucial velocity for droplet formation can be expressed in terms of the Weber number, $W_e = \frac{\rho_l r v_l^2}{\sigma} > 4$, that is

$$v_l > \sqrt{\frac{4\sigma}{\rho_l \cdot r}} \qquad (1)$$

where $r$ is the radius of the jet, $\rho_l$ is the liquid density, and $\sigma$ is the surface tension of liquid jet. The similar process by ejecting gas jet to a soap film is broadly used to generate soap bubbles, and a crucial velocity of gas jet ($v_c$) is found recently [13], which reports that only when gas speed exceeds this threshold $v_c$ can the soap bubbles be formed. When the gas speed $v_g$ is lower than $v_c$, a hemispherical cavity can be formed in the film, and its radius of curvature ($R_c$) decreases with $v_g$. When $R_c$ is equal to the gas jet's radius at the soap film (R(δ)), the soap bubbles are formed, herein δ is the distance from the gas nozzle to the soap film. By comparing the dynamic pressure of the gas jet $\frac{1}{2}\rho_g v_g^2(\delta)$, with the Laplace pressure of the cavity $4\gamma/R_c$, the threshold $v_c$ for bubble generation can be well predicted [13],

$$\frac{1}{2}\rho_g v_c^2(\delta) = 4\gamma/R_c \qquad (2)$$

where γ is the surface tension of the soap solution. When the gas nozzle almost attaches to the soap film ($\delta = 0$), the radius of the gas jet $R(\delta)$ equals to the radius of the nozzle $R_0$, then the velocity threshold can be expressed in Eq. 3.

---

[*] Corresponding Author: hanjf@scu.edu.cn

$$v_c = \sqrt{\frac{8\gamma}{\rho_g R_0}}, \text{ for } \delta = 0 \qquad (3)$$

And when the gas nozzle keeps a certain distance from the soap film ($\delta > 0$), $R(\delta)$ will be bigger than the radius of the nozzle. In the study of Louis Salkin et al. [13], the gas jet was found to follow a conical profile with an opening angle of $23.6°$. In this case, the radius of the gas jet at the soap film can be written as: $R(\delta) = R_0 + \delta \cdot \tan(23.6°/2) \simeq R_0 + \frac{\delta}{5}$, and the gas velocity at the film can be written as $v_c = v_g(\delta) \approx \frac{R_0}{R(\delta)} \cdot v_g$. Then the velocity threshold can be given by Eq. 4 [13].

$$v_c = \sqrt{\frac{8\gamma}{\rho_g R_0}}(1 + \frac{\delta}{5R_0}), \text{ for } \delta > 0 \qquad (4)$$

It can be seen that the difference between the two velocity thresholds $v_c$ for $\delta = 0$ and $\delta > 0$ is mainly caused by the profile of the gas jet which simultaneously influences the key parameters for the bubble generation such as the dynamic pressure $\frac{1}{2}\rho_g v_g^2(\delta)$ and Laplace pressure $4\gamma/R_c$.

In this work, one long-life vertical two-dimensional soap film similar to that described in reference [14, 15] was established, and soap bubble formation mechanism for $\delta > 0$ was precisely studied. Analogous to the results reported by Louis Salkin [13], our experiment results showed that bigger bubbles were generated when the gas velocity was higher than one threshold, and this threshold (defined as $v_{b\_th}$ in this letter) can be well described by Eq. 4.

However, we found that small bubbles are formed even when the gas jet velocity is lower than the predicted threshold, and this small bubble formation threshold (defined as $v_{s\_th}$ in this letter) can be roughly described by Eq. 3. We also found one velocity range that no bubbles are formed when the gas velocity is higher than $v_{s\_th}$ but lower than $v_{b\_th}$. After detailed study of the bubble formation process, we found the experimental results can be explained very well by the aerodynamics of the gas jet.

One soap liquid circulation system was constructed to generate a long-life soap film, and the bubbles were generated when the gas ejected from the nozzle were blew to the film. The critical factors such as the gas flow density and velocity, the thickness of the film and the distance from the nozzle to the film were controlled precisely to study the formation mechanism of the bubbles.

High purity nitrogen gas was used to generate the gas jet. The inner radius of the nozzle $R_0$ was kept constant at 0.5mm. The pressure and flow rate of the gas jet were carefully controlled to get the required velocity. The gas flow rate $q_g$ was varying within the range of 0.5-2.0 L/min, and the average velocity of the gas jet can be calculated from dividing the gas flow rate by the nozzle area, $v_g = q_g/(\pi R_0^2)$, which ranged from 10.6-42.4 m/s. Hence the Reynolds number of the gas flow $R_e = \rho_g v_g R_0/\eta_g$ was in the range of 350-1800, where $\eta_g$ is the gas dynamic viscosity. The distance $\delta$ between the nozzle and the film was varying in the range of 7-20 mm. In this test, the flow rate of soap solution $q_l$ was kept to 12 mL/min, and the width of the film $w$ was kept to 35mm, then the thickness of the film $e \propto (q_l/w)^{(3/5)}$ [13, 15] was about 5μm, which is thin enough [13] to make the threshold speed independent of the film thickness and the hydrodynamics. The same commercial soap solution was used to keep its viscosity $\eta$ and surface tension $\gamma$ constant. A high-speed digital camera (Photron Fastcam UX50) with a lens (NIKON 24-70mm/F2.8) was used to record the generation process of the soap bubbles, the speed of the camera was fixed to 2000 fps and the shutter time was fixed to 0.5 milliseconds to capture the whole process. And one strong light source was used to obtain good brightness images.

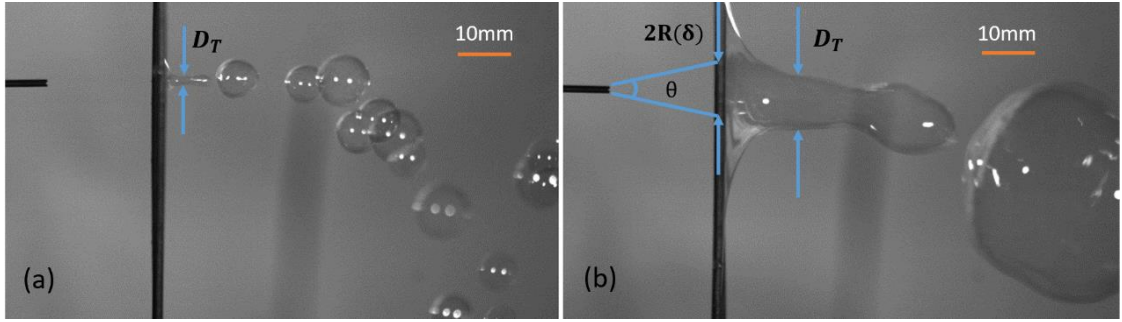

**FIG.1 Snapshots of the formation of one small and big soap bubble respectively at $v_g = 16.3\text{m/s}$ and $v_g = 42.4\text{m/s}$ where δ is fixed at 19.5mm. The average diameter of the tube cavity is defined as $D_T$.**

Normally one hemispherical cavity would be generated when the gas jet hits the soap film. When the gas jet velocity exceeds the threshold, the cavity is elongated heavily and one soap bubble is formed at the end of the cavity. The typical images when one small and big soap bubble is just formed are shown in Fig.1 (a) and Fig.1 (b), and the gas velocities are higher than $v_{s\_th}$, $v_{b\_th}$, respectively.

When δ is fixed to 19.5mm, the bubble formation process for gas jet velocity $v_g$ ranging from 0 to 45 m/s are shown in Fig.2. The process in Fig.2 can be divided into four phases.

For phase I (image a), when the gas velocity is lower than 16.3 m/s, the dynamic pressure inside the cavity is less than the needed pressure to generate bubbles. In this case, only one small cavity is generated in the soap film and bubbles cannot be formed. The cavity's radius of curvature $R_c$ decreases with the gas velocity, which is consistent with the phenomenon of Louis Salkin's study [13].

For phase II (image b1-b7), when the gas velocity is higher than 16.3 m/s but lower than 27.6 m/s, the cavity is elongated to a long cylindrical tube and small bubbles are generated at the end of the cavity. These two velocity thresholds are defined as $v_{s\_th}$ and $v_{s\_up}$, which means small bubbles are generated when the gas velocities are within this range. Herein, it is found that $v_{s\_th}$ is much smaller than the predicted threshold $v_c(\delta = 19.5mm)$ in Eq. 4, but is almost equal to the threshold $v_c(\delta = 0)$ in Eq. 3.

For phase III (image c1-c4), when the gas velocity is higher than 27.6 m/s but smaller than 42.4 m/s, soap bubbles cannot be formed and one much bigger cavity is generated. It is also found the curvature radius of the cavity decreases with the gas velocity, but its size is much bigger than that in phase I.

For phase IV (image d), when the gas velocity is higher than 42.4 m/s, one larger cylindrical cavity is elongated and big bubbles are formed at the end of the cavity. This velocity threshold is defined as $v_{b\_th}$, which is consistent with the predicted $v_c(\delta = 19.5mm)$ in Eq.4.

In summary, the bubbles can be formed when the gas velocity is inside two ranges, small bubbles are formed for the small velocity range of ($v_{s\_th} = 16.3m/s$, $v_{s\_up} = 27.6m/s$), and big bubbles are formed when the velocity is higher than $v_{b\_th}$ ($42.4m/s$). The experiment results demonstrate that the bubble formation mechanism is more complicated than that reported in reference [13] which said that the bubbles were formed when the gas blowing velocity was above one threshold.

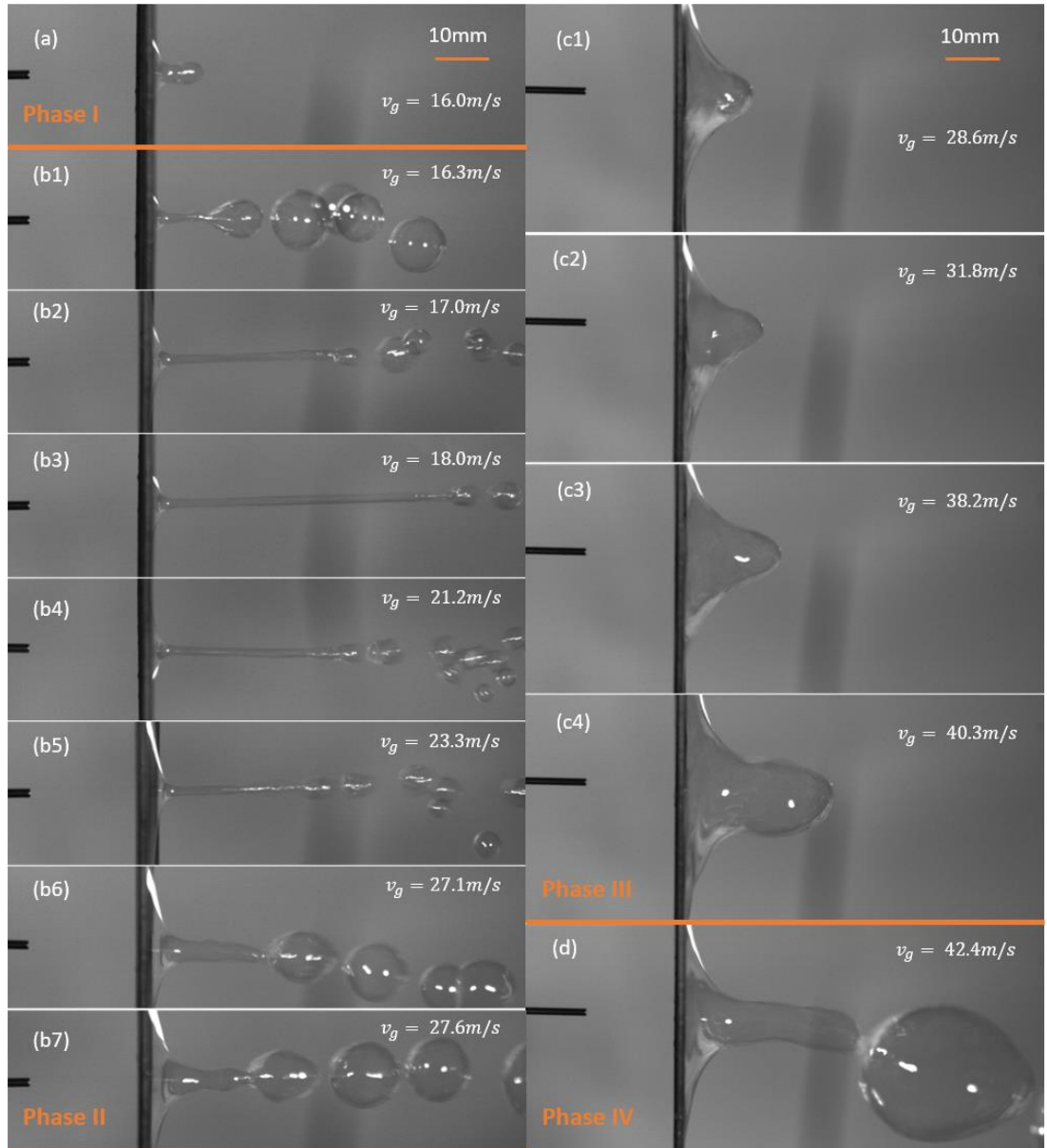

**FIG.2 Snapshots of soap bubbles formation at different gas velocities when δ is fixed to 19.5 mm. Small bubbles can be formed when the velocities are within the range of (16.3m/s, 27.6m/s), big bubbles can be formed when the gas velocity is higher than 42.4m/s, and no bubbles are formed for the velocity range of (27.6m/s, 42.4m/s).**

In order to study the influence of the distance δ between the nozzle and the soap film, the same processes in Fig.2 were tested for other four distances (δ =7.5mm, 10.5mm, 13.5mnm and 16.5mm). The similar four phases in Fig.2 were observed when δ is greater than or equal to 10.5mm, while only one velocity threshold was observed for the δ = 7.5mm case, which are much different from Louis Salkin's study [12]. To ascertain the reason of this discrepancy, an in-depth analysis has been made to explain the results. In Louis Salkin's study [13], the gas jet was treated as turbulent flow when the Reynolds number varied from 500-5000. However, we believe the gas jet with Reynolds number $R_e$ ranging from 350-1800 in this work should be treated as laminar flow for lower velocities $(R_e < 1000)$ and might transform into turbulent flow for higher velocities $(R_e > 1200)$. And it can be inferred that the aerodynamics of the gas jet may be the main reason of the velocity threshold discrepancy.

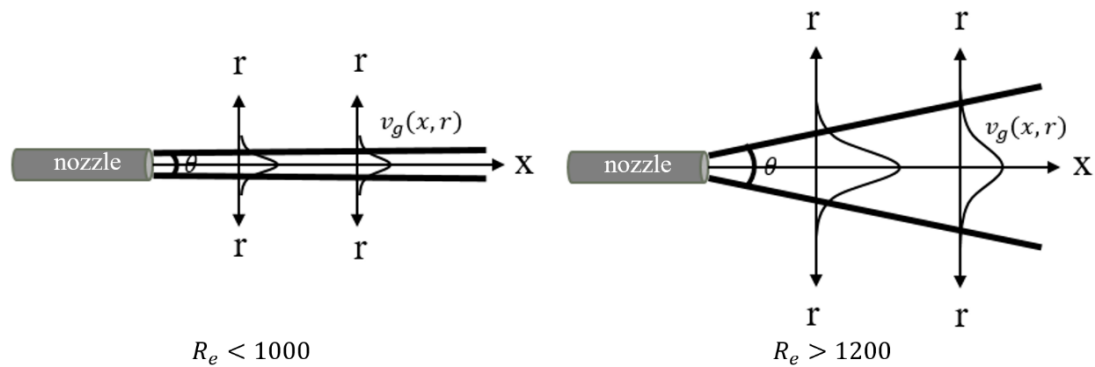

**FIG.3 Schematic of profiles of the gas jet ejecting from one cylindrical nozzle into the quiescent air for low Reynolds number and high Reynolds number. It is found the opening angel is $0.9°$ when the Reynolds number is lower than 1000, and is $23.6°$ when the Reynolds number is higher than 1200 in our test.**

The laminar and turbulent free jet has been studied widely both by experiment and theory [16-19]. It was reported in reference [20] that the pressured air ejecting from a nozzle into a quiescent air adopted a conical angle of 23.6 °for turbulent gas jet with large Reynolds number. But the opening angle and dissipation of the gas jet for lower Reynolds number was very different from that of higher Reynolds number [21, 22]. The schematic profiles of laminar flow and turbulent flow are shown in Fig.3. The opening angle and velocity dissipation of gas jet with small Reynolds number are much smaller than that with large Reynolds number. Then it can be inferred that gas jets with different Reynolds numbers might have different opening angles, in other words, the opening angle $\theta$ might change with the gas velocity $v_g$ which is proportional to Reynolds numbers.

For the gas jet generated from one cylindrical nozzle with a conical opening angle, the radius of the gas jet at distance $\delta$ satisfies $R(\delta) = R_0 + \delta \cdot \tan\frac{\theta}{2}$. It is found in Fig.1 and Fig.2 that once the cavity is elongated into a tube, the cavity will maintain the same diameter. Then we can suppose that the diameter of the cylindrical cavity $D_T$ equals to the diameter of the gas jet at the soap film $2R(\delta)$, which can be written as Eq. 5.

$$D_T = 2R(\delta) = 2(R_0 + \delta \cdot \tan\frac{\theta}{2}) \quad (5)$$

It can be inferred from Eq. 5 that for one given opening angle $\theta$ (which is connected with a certain gas velocity $v_g$), the diameter of the cavity will increase linearly with $\delta$. The diameters of the cavity $D_T$ at two velocities are shown in Fig. 5(a), which can fit with Eq. 5 very well. And $\theta$ can be calculated from the fitted slope. The opening angles for other gas velocities are calculated and the results are shown in Fig. 5(b). It is found in Fig. 5(b) that the curve can be divided into three parts. For the gas velocities lower than 25.5 m/s (Re<1000), the opening angle keeps constant at $0.9°$, and the gas jet can be treated as laminar flow. When the gas velocity is higher than 29.5 m/s (Re>1200), the opening angle keeps constant at $21.6°$, and the gas jet can be treated as turbulent flow. It must be pointed that the fitted angle of $21.6°$ for turbulent flow is slightly smaller than the $23.6°$ of air free jet, which might be caused by the geometry of the nozzle as well as the gas type. When the Reynolds number increases from 1000 to 1200, the opening angle increases from $0.9°$ to $21.6°$, and the gas jet can be considered as transition flow. The opening angles for this transition zone are fitted linearly, and the fitted function is shown in Eq. 6.

$$\theta = 3.6 \cdot v_g - 92 \tag{6}$$

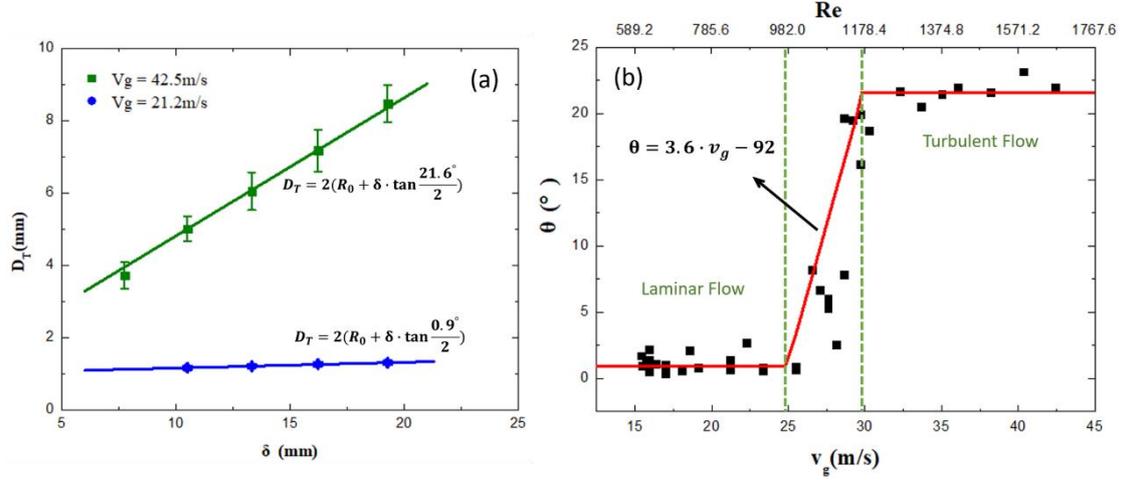

**FIG.4 (a). Plots of cavity diameter $D_T$ vs $\delta$ (distance from gas nozzle to the soap film) for gas speed $v_g = 21.2 \text{m/s}$ and 42.5m/s, the curves are fitted with Eq. 5 and the fitted θ are $0.9°$ and $21.6°$ respectively. (b). The opening angle of the gas jet (θ) at different gas velocity and Reynolds number, the curve can be divided into three parts which are connected to laminar flow, transition flow and turbulent flow.**

Then the velocity threshold for soap bubble generation can be discussed into three cases.

Case I, for the laminar flow (Re <1000), the opening angle of the gas jet is $0.9°$, and the gas speed is below 25.5 m/s in our experiments. The radius of the gas jet at the soap film is $R(\delta) = R_0 + \delta \cdot \tan(0.9°/2)$ and the average gas speed at the soap film is $v_g(\delta) = \frac{R_0}{R(\delta)} \cdot v_g$. Thus, the threshold speed $v_{s\_th}$ can be calculated based on Eq. 2, which is described as Eq. 7:

$$v_{s\_th} = \sqrt{\frac{8\gamma}{\rho_g R_0}\left(1 + \frac{0.008 \cdot \delta}{R_0}\right)} \tag{7}$$

This predict velocity threshold for small bubble generation is depicted in Fig. 6 (green solid line), we can see the tested data $v_{s\_th}$ follows the formula pretty well. Since the opening angle of the gas jet is very small for the laminar flow, the distance between the nozzle and the soap film has little effect on the velocity threshold. And this is why $v_{s\_th}$ obtained in our test increases slightly with distance $\delta$ and is very close to $v_c(\delta = 0)$ in Eq.3.

Case II, for the turbulent flow (Re > 1200), the opening angle of gas jet keeps constant at about $21.6°$, and the gas speed $v_g$ is greater than 29.5m/s. The radius of the gas jet at the soap film is $R(\delta) = R_0 + 0.19 \cdot \delta$ and the average gas speed at the soap film is $v_g(\delta) = \frac{R_0}{R(\delta)} \cdot v_g$. Therefore, the threshold speed $v_{b\_th}$ can be calculated based on Eq. 2, which is expressed as:

$$v_{b\_th} = \sqrt{\frac{8\gamma}{\rho_g R_0}\left(1 + \frac{0.19 \cdot \delta}{R_0}\right)} \tag{8}$$

The tested velocity thresholds of big bubble generation $v_{b\_th}$ are in well accordance with Eq. 8 as shown in Fig. 6. Since the fitted opening angle is $21.6°$, which is slightly smaller than the $23.6°$ reported in reference [13], the tested threshold for our experiments is slightly smaller.

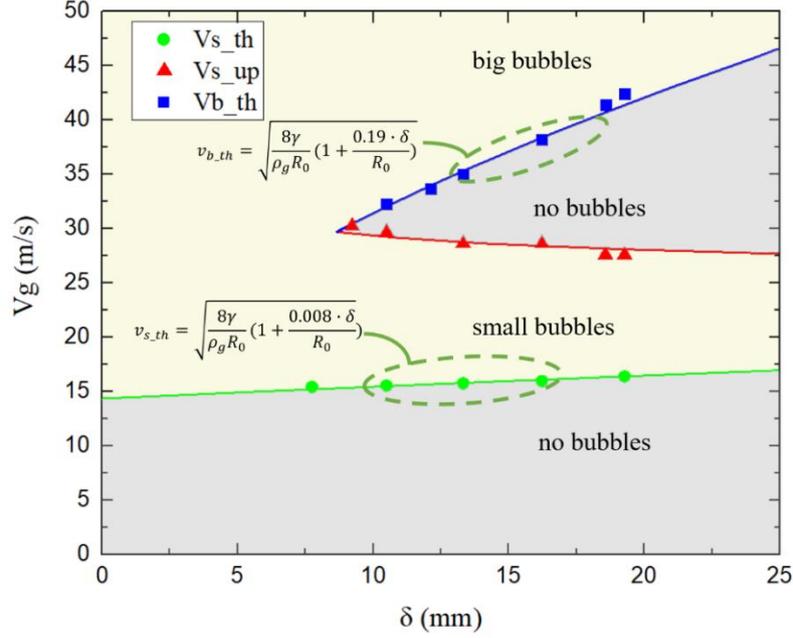

**FIG.5 The gas velocity threshold for soap bubble generation. When $\delta > 8mm$, Small bubbles are formed for the small velocity range of ($v_{s\_th}$, $v_{s\_up}$), and big bubbles are formed when the velocity is higher than $v_{b\_th}$. When $\delta < 8mm$, the soap bubble formation has only one threshold $v_{s\_th}$.**

Case III, for the transition zone ($1000 < R_e < 1200$), the opening angle of gas jet increases linearly from $0.9°$ to $21.6°$, and the gas speed $v_g$ is in the range of 25.5m/s-29.5m/s. The radius of the gas jet at the soap film is $R(\delta) = R_0 + \delta \cdot \tan\frac{3.6v_g - 92}{2}$, and the average velocity at the soap film is $v_g(\delta) = \frac{R_0}{R(\delta)} \cdot v_g$. Therefore, the threshold speed $v_c$ can be calculated based on Eq. 2, which is written as:

$$v_c = \sqrt{\frac{8\gamma}{\rho_g R_0}(1 + \frac{\delta \cdot \tan((3.6v_c - 92)/2)}{R_0})} \tag{9}$$

Though it is not easy to give the exact formula to calculate the velocity threshold for the transition range, the threshold value for each $\delta$ can be calculated from Eq. 9. The calculated $v_c$ and the tested $v_{s\_up}$ are shown in Fig.5, and they can fit each other very well. Then it can be inferred that the transition of the gas jet from laminar flow to turbulent flow will cause one velocity range that no bubbles can be formed as shown in Fig. 5. But this is only effective when $\delta$ is bigger than 8mm. When $\delta$ is smaller than 8mm, the big cavity in phase III will not appear and soap bubbles are always generated. The calculated threshold $v_{s\_up}$ in Eq. 9 is always bigger than the threshold $v_{b\_th}$ calculated from Eq. 8, which means big bubbles can be generated and Eq. 9 cannot be used for this case. The reason is that when the distance between the nozzle and the soap film is small, the gas jet cannot expand broadly even for turbulent flow, and only one velocity threshold for laminar flow is available for this case.

In conclusion, the soap bubble formation is one competition between the inertia force of the gas jet and the surface tension of the soap film, and the aerodynamics of the gas jet in ambient air has a significant influence on the soap bubble generation. When the distance between the nozzle and the soap film is big (bigger than 8mm in our experiments), there are two gas velocity ranges for the soap bubble generation, small bubbles will be generated for the small gas velocity range of ($v_{s\_th}$, $v_{s\_up}$), and big bubbles will be generated when the gas velocity is higher than $v_{b\_th}$. The small bubble velocity threshold of $v_{s\_th}$ is connected with the laminar gas jet with a lower Reynolds number, while the big bubble velocity threshold of $v_{b\_th}$ is connected with the turbulent gas jet with a bigger Reynolds number. And the transition from laminar flow to turbulent flow will cause one velocity range that no bubbles can be generated. When the distance between the nozzle and the soap film is small(less than 8mm in our experiments), the gas flow type has little influence on the bubble generation process, and bubbles are generated once the gas velocity is higher than the small threshold $v_{s\_th}$.

This work is supported by the National Magnetic Confinement Fusion Program of China (2014GB125004) and National Natural Science Foundation of China (11575121).